%% file: main.tex
\documentclass[
aps,
prb,
twocolumn,
superscriptaddress,
showpacs,
english,
nofootinbib,
floatfix]
{revtex4-2}

\usepackage{mymacros}
\usepackage{graphicx}
\usepackage{placeins}
\usepackage{float}

\begin{document}

\title{Targeting high symmetry in structure predictions by biasing the potential energy surface}

\author{Hannes Huber} \affiliation{\UniBasel}
\author{Martin Sommer} \affiliation{\UniBasel}
\author{Moritz Gubler}\affiliation{\UniBasel}
\author{Stefan Goedecker} \affiliation{\UniBasel}

\begin{abstract}
    Ground state structures found in nature are in many cases of high symmetry. But structure prediction methods typically render only a small fraction of high symmetry structures. Especially for large crystalline unit cells 
    there are many low energy defect structures. For this reason methods have been developed where either preferentially high symmetry structures are used as input or where the whole structural search is done within a certain symmetry group. In both cases it is necessary to specify the correct symmetry group beforehand. However it can in general not be predicted which symmetry group is the correct one leading to the ground state. For this reason we introduce a potential energy biasing scheme that favors symmetry and where it is not necessary to specify any symmetry group beforehand. On this biased potential energy surface, high symmetry structures will be found much faster than on 
    an unbiased surface and independently of the symmetry group to which they belong. 
    For our two test cases, a $C_{60}$ fullerene and bulk silicon carbide, 
    we get a speedups of 25 and 63. 
    In our data we also find a clear correlation between the similarity of the atomic environments and the energy. In low energy structures all the atoms of a species tend to have similar environments. 
\end{abstract}

\maketitle

Structure prediction methods are an important tool for  the discovery of new materials~\cite{Oganov.2019}. Such methods can not only be applied for materials at ambient pressure but also under very high pressures that are relevant for geophysical applications, but not accessible by experimental methods~\cite{zurek}. For this reason, numerous methods such as  
simulated annealing~\cite{anneal}, basin hopping~\cite{basin_hopping}, minima hopping  (MH)~\cite{mh,Amsler2010,Sicher2011,Roy2008}, random structure searches~\cite{Pickard2006}, meta-dynamics~\cite{martonak} and various variants of evolutionary genetic algorithms~\cite{Johnston.2003, Oganov.2006, Bhattacharya(2013), Vilhelmsen.2014, Zhu.2015, Curtis.2018} as implemented in the USPEX~\cite{USPEX}, CALYPSO~\cite{calypso} and XtalOpt~\cite{Xtalopt} software package have been developed. These advanced global geometry optimisation methods have shown that they can efficiently explore\cite{Jrgensen.2018, Yamashita.2018} the potential energy surface (PES) of different clusters and bulk materials under a variety of external conditions and predict new structures. 
However, such methods require a high computational effort because the number of possible meta-stable structures grows exponentially with respect to the number of atoms in the system and the generation  and relaxation of a single structure requires many energy and force evaluations.

Unless defects of certain materials are studied explicitly,
the ground state and the lowest defect free meta-stable structures are of greatest interest since they are the structures that can most likely be synthesized. 
However, for large cells most structures that are found in a structure prediction contain defects. Typically these defect structures represent local minima in a funnel whose bottom corresponds to a defect-free, meta-stable or the global minimum structure.
To find this relatively small number of defect-free structures, a structure search which visits a very large number of defect structures, is inefficient.\\
To favor high symmetry, most crystal structure prediction methods use input guess structures that are of high symmetry. If the correct symmetry is chosen, the most similar low energy structure is found much more rapidly. For basin hopping and genetic algorithms there exist also a versions where all the moves of the atoms are constrained to conserve the desired symmetry~\cite{SASS,basin_hopping}. The inconvenience in all these approaches is that there are more than 200 space groups and it is a priori unknown which one will be adopted by the system. Traditionally symmetry is defined by geometric operations such as rotations or reflections that leave the structure invariant. 
We will use in this work an alternative definition of symmetry.
We will consider a system to be highly symmetric if all atoms of the same element see only a small number of different environments. 
Structures with a large number of environments are actually unlikely to exist 
according Pauling's rule of structural parsimony originally established for ionic materials~\cite{Pauling.1929, Pauling.1989}.
In many cases we will actually try to find systems where all atoms of 
the same element see the same environment. Evidently this is true for many 
high symmetry structures such as the $C_{60}$ fullerene or the diamond structure of silicon and carbon. 
A structure is either invariant or not under certain symmetry operations. 
So basing some penalty function on the number of possible symmetry operations 
would give rise to a discontinuous function. Our definition of similarity is 
however a continuous functions. It is zero if the environments are identical and grows larger in a continuous way when the environments become more different. 
Our definition is thus broader than the traditional one.
We classify a structure also as highly symmetric if there are a few distinct environments which are however very similar.

The tendency that low energy structures have in general similar environments has already been exploited to gain efficiency in the context of evolutionary structure prediction algorithms. In this context mutation moves were introduced that favor environments that have a similar radial distribution as certain selected role model environments~\cite{ClusterReg,ConvexPES}. 
Even in amorphous systems it was observed that structures that had similar pair distribution functions were also low in energy~\cite{amorphous}. 

The basic idea of our approach is to perform a structure search on a biased potential energy surface ~\cite{De.2019}
$$
E_b(R_1,\ldots,R_n)=E(R_1,\ldots,R_n) + \omega P(R_1,\ldots,R_n)
$$
where $E_b$ is the biased PES, $E$ is the physical PES, $P$ is the penalty function and $\omega$ is the biasing weight.
Since the number of environments is larger for defective structures, the penalty part will push up these defective structures on the biased PES. In this way the 
downhill barriers are lowered compared to the uphill barriers and the 
PES becomes a stronger structure seeker character which speeds up the search 
for the global minimum and possibly other high symmetry structures at the bottom of other funnels.

We quantify the similarity of environments with the overlap matrix (OM) fingerprint~\cite{zhu2016fingerprint} based on s- and p-type orbitals, that was shown to be able to detect in a highly reliable 
way different atomic environments~\cite{Parsaeifard.2020}. 
In particular this fingerprint has both radial and angular resolution. In the OM method the 
eigenvalues of a localized overlap matrix, centered on the atom $i$ whose environment has to be characterized, are assembled into an atomic environment fingerprint  vector   $ \bf{f}_i$. This environment characterization is done for all atoms in the system.
If all atomic environments are identical, the rank of the matrix $F$ formed by all these  
vectors $\bf{f}_i$ is one, if there are two distinct elemental environments the rank is two, etc. The rank can most easily be calculated from the eigenvalues $\lambda_i$ of the Gram matrix $D=F^TF$, constructed from the these fingerprint vectors. 
The number of the non-zero eigenvalues of this matrix gives the rank of the fingerprint vectors.
So the penalty function that favours one single environment for a certain element is
\begin{equation}
P_1(R_1,\ldots,R_{N_{at}})=\sum_{i=2}^{N_{at}} \lambda_i=\Tr(D) -\lambda_1,
\end{equation}
In case we want to allow for up to two environments, the penalty becomes 
\begin{equation} 
P_2(R_1,\ldots,R_{N_{at}})=\sum_{i=3}^{N_{at}} \lambda_i=\Tr(D) -\lambda_1 -\lambda_2,
\end{equation}
where $\Tr$ is the trace of the matrix, i.e. the sum over all eigenvalues. 
As usual, we have assumed in all the above formulas that the eigenvalues are sorted in decreasing order. 
For a multi-component system, each element contributes its own  
penalty function and the total penalty function is the sum of all the elemental contributions.
For highly symmetric structures where all local environments are equivalent, e.g. the ground state of C$_{60}$, the bias function will be exactly zero. If the environments get more distinct, the bias function grows due to the positive semi-definiteness of the Gram matrix. Since the Gram matrix gives essentially the effective dimension of the vector space spanned by the local descriptor vectors, it is called dimensionality matrix in this paper.

To test our method we selected two systems of quite different nature. The first one, silicon carbide, is a crystalline system with two elements that have to mix in the right way to find low energy structures and the second, the 
C${60}$ fullerene, is a molecular clusters. Its global minimum is just one structure out of a huge number of meta-stable structures with varying structural motifs such as planar structures, chains and bowls.

For the exploration of the PES the minima hopping (MH) algorithm was used, but our biasing scheme is in principle applicable to any structure prediction method. The MH algorithm is not based on thermodynamic principles like simulated annealing or basin hopping but uses a combination of molecular dynamics, local geometry optimization and a history of previously found local minima to escape quickly from already known regions and hence, efficiently explore the entire PES. For the geometry optimization of C$_{60}$ with free boundary conditions the  conjugate gradient method was used. In the case of periodic boundary conditions (PBC) the highly efficient and stable vc-SQNM method developed by Gubler et al.~\cite{moritz_SQNM, Schaefer.2015} was used. 
The molecular dynamics (MD) simulation was implemented using the velocity Verlet algorithm for the non-periodic case and the variable cell shape MD~\cite{parrinello-md} for PBC. This method allows atoms as well as cell vectors to move dynamically during the MD simulation for PBC.\\
For the calculation of the PES of C$_{60}$ the transferable tight binding potential for carbon from Xu et al.
~\cite{xu1994tight} was used. For the silicon carbide simulations in PBC DFTB+~\cite{Hourahine.2020.DFTB+} was used with the Slater-Koster parameterisation set pbc-0-3~\cite{Sieck.pbc-0-3}. 
The MD and an initial local geometry optimisation are performed on the biased PES followed by a local geometry optimisation on the unbiased PES.
This avoids falling in potentially existing spurious local minima on the biased PES.

To obtain conservative forces of the biased PES the derivative of the symmetry bias needs to be added to the physical forces. The same is true for the derivative of the biased symmetry function with respect to the lattice vectors which need to be added to the lattice derivatives in the case of PBC. The derivations of these two quantities can be found in the supplementary information. \\
\begin{figure}[htb]
    \includegraphics[width=8.6cm]{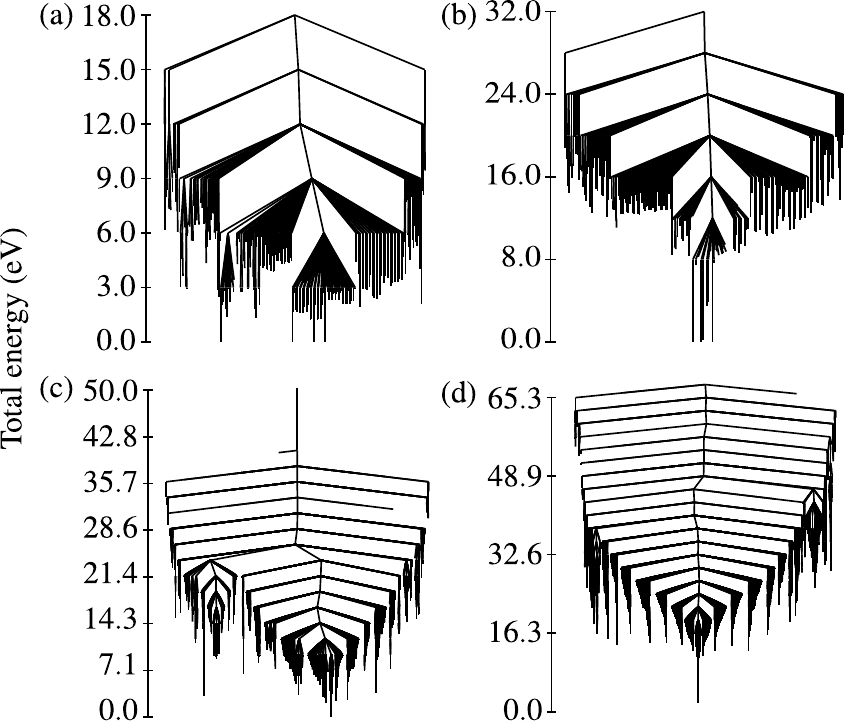}
    \caption{Changes in the characteristics of the disconnectivity graphs of silicon carbide (top row) and $C_{60}$ (bottom row) induced by a bias. The left column shows the disconnectivity graphs of the PES without a bias and the right column   the disconnectivity graphs of the PES with a bias.
    The graph was constructed with the disconnectionDPS software~\cite{dpsSoft}.
    }
    \label{fig:merged_small}
\end{figure}

The character of a PES can best be deduced from the appearance 
of its disconnectivity graph~\cite{disconnect} Fig. \ref{fig:merged_small}. For a structure seeker~\cite{walesbook}, the downhill barriers are much lower than the uphill barriers. As a consequence, any algorithm that 
crosses preferentially lower barriers will experience some driving force 
toward the minimum at the bottom of the funnel and find it therefore faster. This driving force will of course depend on the strength $\omega$ of the bias. While on the one hand it is desirable to choose a large $\omega$, the penalty should on the other hand only induce some 
weak perturbation that does not completely deform the physical PES.
In particular there should remain in most cases a one-to-one mapping between the local minima on the physical and the biased PES. As already noted by 
Zwanzig in the context of protein folding~\cite{zwanzig} a relatively small 
bias can have a large effect on the dynamics of the system and reduce the folding time by several orders of magnitude. We were indeed 
always able to find a range of values for $\omega$ that did speed up the 
search for high symmetry structures considerable without destroying the overall character of the PES. With our weight the downhill barriers are typically twice as large as the uphill barriers and the penalty difference between high and low symmetry structures 
is a few times the difference of their physical energy. This later criterion can be used to find suitable values of $\omega$. 

Fig.~\ref{fig:merged_small} shows the differences of the disconnectivity graphs 
for the unbiased and biased system.
The changes in the appearance of the disconnectivity graphs indicate that the biased PES has a much stronger structure seeker character which should 
make the search for the lowest high symmetry structures considerably faster.

\begin{figure}[htb]
    \includegraphics[width=8.6cm]{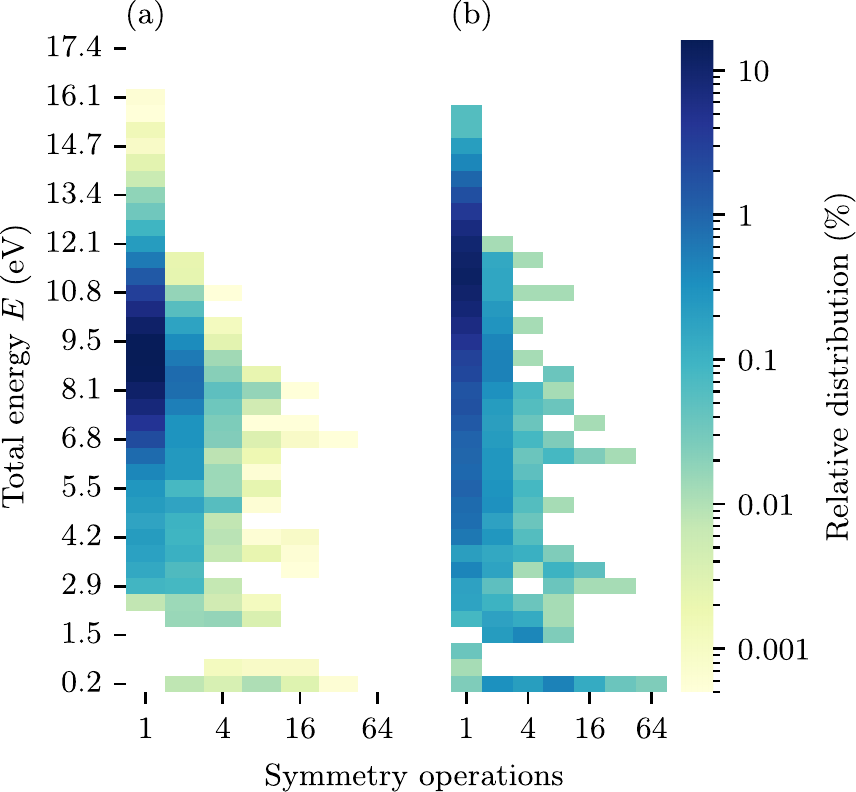}
    \caption{Comparison of the distribution of the found minima on the unbiased (a) and biased (b) PES. The symmetry, plotted along the x-axis, is measured by the number symmetry operations that 
    leave the structure invariant. The coloring indicates the relative abundance of 
    meta-stable structures for a given symmetry and energy.}
    \label{fig:heatmap}
\end{figure}
\begin{figure}[htb]
\includegraphics[width=8.6cm]{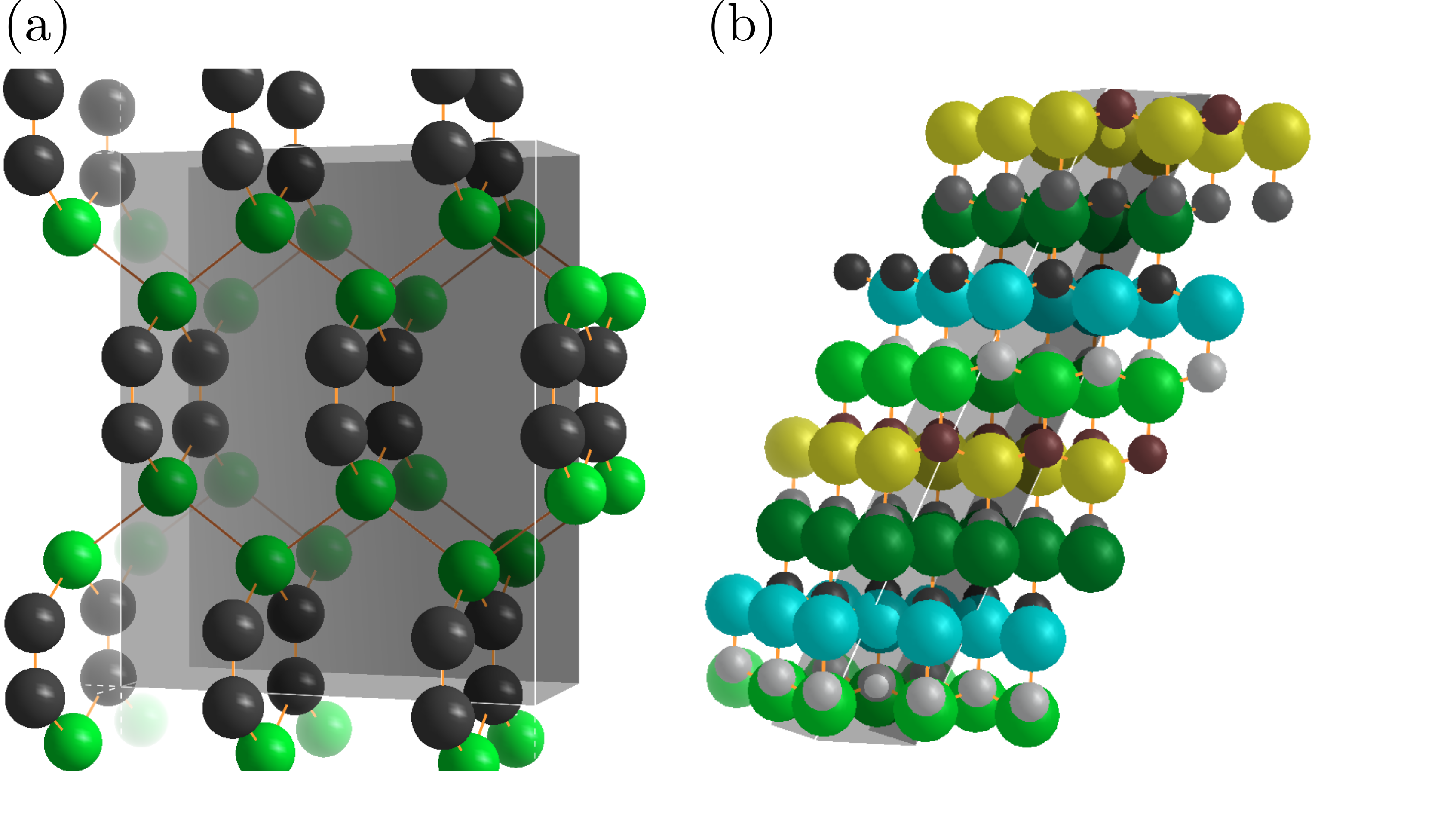}
\caption{Two high symmetry structures found by a biased run. The structure (a) is only 517 meV/atom higher in energy than the ground state structure, even though the bonding character is completely different from the ground state where all atoms are 4-fold coordinated. In this structure all carbon atoms are 3-fold coordinated. Structure (b) is a 4-fold coordinated silicon carbide structure with an energy of 6 meV/atom above the ground state. Four different environments exist for each carbon and silicon atom, but the environments are so similar that the differences can not be detected by eye. Carbon atoms are displayed by smaller spheres than silicon atoms for which each environment has its own colour.
} 
\label{fig:3fCstructure}
\end{figure}
\begin{figure}[htb]
    \includegraphics[width=8.6cm]{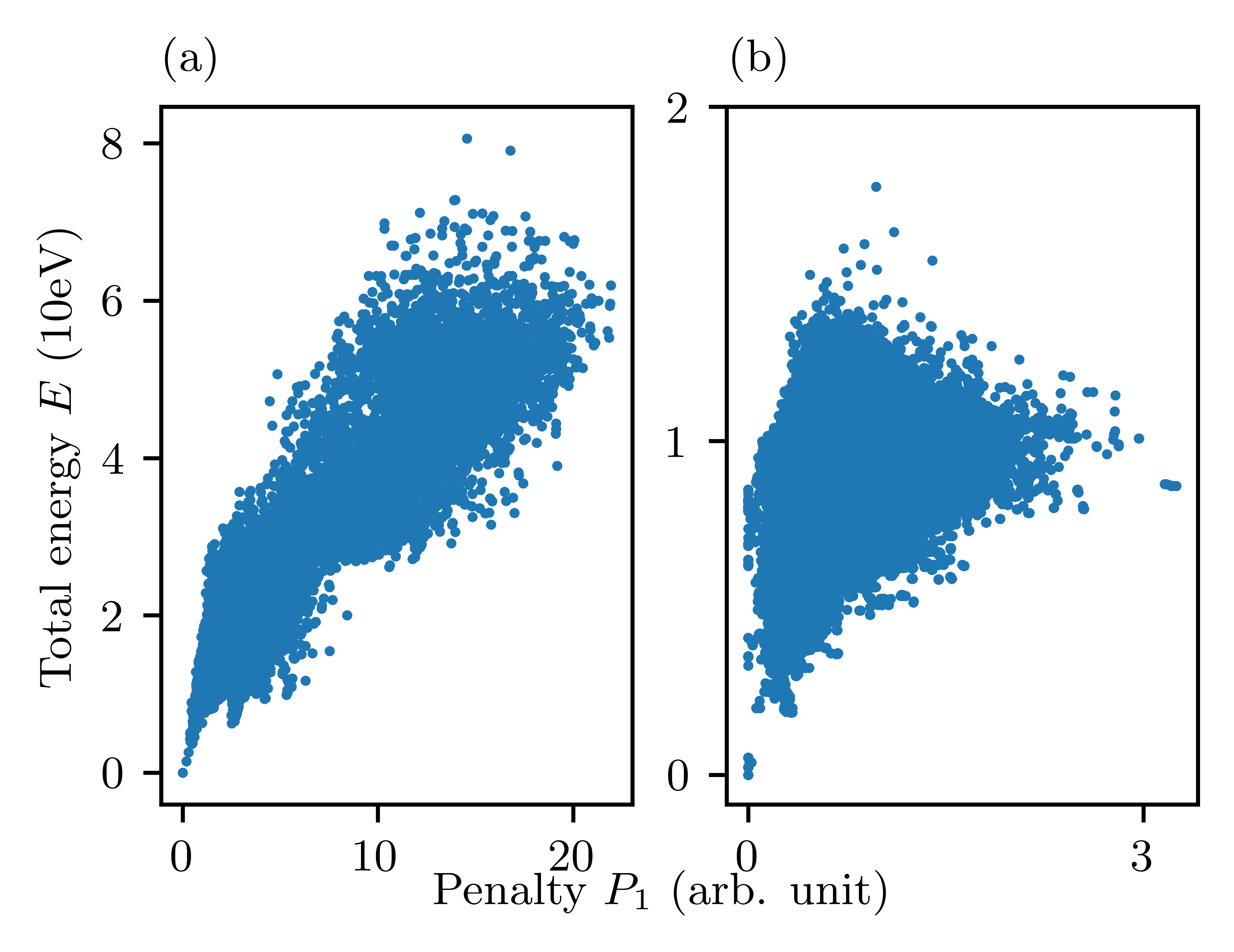}
    \caption{Correlation between our penalty function $P_1$, that is a measure of the structural diversity and the total energy $E$. 
    Only if all atoms in the C$_{60}$ (a) or silicon carbide (b) structure have similar environments, the energy of the structure will be low. }
    \label{fig:plt_scatter_nb_SiC}
\end{figure}

To investigate the effect of the symmetry bias on the speed of the global geometry optimization in a systematic way, statistical tests were conducted for $C_{60}$ and 16 atom silicon carbide cells. One hundred global geometry optimizations were started from different initial configurations until the ground state structure, or in the case of silicon carbide a polytype of the ground state, was found on the unbiased and the biased PES. 
For $C_{60}$ the carbon atoms were randomly placed on a plane and for silicon carbide the carbon and silicon atoms were randomly placed in spatially separated sub-cells that formed the crystalline cell. 
To avoid nonphysical structures a minimum and maximum distance between the randomly placed atoms was enforced.
Since there are no phase separated low energy structures in a cell of this size, the silicon and carbon atoms always had to mix to find the low energy structures. 
As a measure for the computational cost of the runs we used the number of required local geometry optimizations.
As can be seen from Table~\ref{tbl:all_results} the biasing reduces the average number of geometry optimizations by a 
factor of 25 for C$_{60}$ and by a factor of 63 for silicon carbide. 
It can also be seen from Table~\ref{tbl:all_results} that other statistical markers like the standard deviation (std),  quantiles and the number of geometry optimizations for 
the fastest as well as the slowest simulations decreased by about the same magnitude.
\begin{table}[h]
\small
\caption{Table of statistical markers and biasing parameters of 100 global geometry optimisations started from randomly generated structures for C$_{60}$ and 16 atom silicon carbide cells on the unbiased PES as well as on the biased PES.  
  The statistical markers always relate to the required number of local geometry optimisations. The biasing parameters for C$_{60}$ were $\omega=0.3$ with $\sigma_c=6.0$~\cite{zhu2016fingerprint} and for silicon carbide $\omega=3.5$ with $\sigma_c=4.5$. 
  The numbers in parenthesis give the speedup with respect to the unbiased runs for corresponding quantities.
  All simulations were successfully carried out until the ground state structure or in the case of 16 atom silicon carbide a polytype of the ground state structure, was found.
  }
  \begin{tabular*}{0.48\textwidth}{@{\extracolsep{\fill}}lllll}
    \hline
      & C60 unbiased & C60 biased & SiC unbiased & SiC biased\\
    \hline
    mean & 9254.71 & 370.87 (25) & 9715.36 & 153.84 (63) \\
    std & 7716.87 & 246.74 (31) & 10181.81 & 97.92 (104) \\
    min & 1184 & 89 (13) & 771 & 31 (25) \\
    25\% & 4048 & 180 (22) & 3506 & 87 (40) \\
    50\% & 6994 & 309 (23) & 5720 & 122 (47) \\
    75\% & 11765 & 520 (23) & 12170 & 202 (60) \\
    max & 44136 & 1165 (38) & 56213 & 571 (98)  \\
    \hline
  \end{tabular*}
  
\label{tbl:all_results}
\end{table} 
As expected and as shown in Fig.~\ref{fig:heatmap} the distribution of the found structures with respect to their physical energy $E$ and their degree of symmetry is also quite different. 
For the biased MH runs, the fraction of high symmetry structures is considerably higher and the average physical energy of low symmetry structures is higher since many low energy defects were not found.
Many of these high symmetry structures are quite interesting. Searching for structures where all atoms of a certain species have the same environment, we found for instance several SiC structures where all the carbon 
atoms were 3-fold coordinated, whereas all the silicon atoms are 4-fold coordinated. Such a structure 
is shown in Fig.~\ref{fig:3fCstructure}. 
Since our penalty function goes smoothly to zero when the environments get more similar, we actually also found most low energy structures of silicon carbide with up to 4 different environments per atom with a penalty function that favours a single environment. 
It turned out that in these cases, the environments tend to be quite similar and result thus in a small instead of a strictly zero penalty function (see Fig.~\ref{fig:3fCstructure}).
This finding is related to a strong correlation between the structural environment diversity as measured by our penalty function and the total energy as shown in Fig.~\ref{fig:plt_scatter_nb_SiC}. 
\\ 
In summary, based on a non-conventional measure of symmetry, that is motivated by Paulings rule of parsimony, we construct a penalty function that measures the dissimilarity  
between different atomic environments in a structure. 
To construct the penalty function no guesses 
of which symmetry will be adopted by the system are required. 
Adding this penalty function 
to the physical PES gives a biased PES where disordered structures are pushed up in energy relative to high symmetry structures. 
This leads to a lowering of the downhill barriers compared to the uphill barriers. 
This stronger structure seeker property of the biased potential energy surface allows for much faster searches for high symmetry ground states.\\
The penalty function also allows us to find high symmetry structures of higher energy rapidly.
This feature opens the way to perform structure prediction on a PES that was constructed with a cheap but not very accurate method to find high symmetry structures in low as well as moderately higher energy regions. 
These high symmetry structures can then be 
reranked by calculating their energies with 
a more accurate but also more expensive electronic structure method. In this way high energy structures that were higher in energy with the cheap method can become low in energy with the accurate method. This procedure would not be possible without the bias because in this case the overwhelming majority of higher energy structures are typically all defect structures, which are unlikely to become low energy structures when reranked. \\
Our results also clearly show the general validity of Paulings rule that in low energy structures 
the variability of the atomic environments is quite limited.\\
Financial support from SNF and computing time from CSCS ( project s963) 
and sciCORE (\url{http://scicore.unibas.ch/}) are acknowledged. 
We thank Prof. Alireza Ghasemi and Prof. Andris Gulans for useful comments on the manuscript.
\pagebreak

\pagebreak
\appendix{}\label{seq:Supplementary information}
\begin{widetext}
\input{appendix}

\end{widetext}

\end{document}

%% file: appendix.tex
\section{Symmetry Bias}
The goal of the symmetry bias is to find a measure for the structural symmetry of the system and to use it as a bias on the PES to drive the system faster to the global minimum during a minima hopping simulation. As a measure for the structural symmetry of a system we quantify the differences between the individual atomic environments. 
\\
First a matrix $F$ is formed containing the overlap matrix (OM) fingerprints ~\cite{zhu2016fingerprint} of every atom $k$ of the cluster or cell as vector $V_k$ in columns. 
In the OM fingerprint method the eigenvalues of a localized overlap matrix are assembled into a vector.
All entries of each fingerprint vectors $V_k$ need to be sorted before forming the matrix $F$.
$$
F=
    \left(
    \begin{array}{ccccc}
      V_1(1) & V_2(1) & V_3(1) & \cdots & V_{nat}(1) \\
      V_1(2) & V_2(2) & V_3(2) & \cdots & V_{nat}(2) \\
      \vdots & \vdots & \vdots & \ddots & \vdots \\     
      V_1(l_{fp}) & V_2(l_{fp}) & V_3(l_{fp}) & \cdots & V_{nat}(l_{fp}) \\
    \end{array}
    \right)
$$
with $V_k(j)$ being the j-th entry of the OM fingerprint and $l_{fp}$ is the length of the fingerprint vectors. 

The Gram matrix 
\begin{equation} 
D=F^TF
\end{equation}
can now be formed.

If all atomic environments are identical, the rank of the matrix $D$ formed by the OM fingerprint vectors is one, when there are only two different environments the rank is two, etc. The rank can most easily be calculated from the eigenvalues $\lambda_i$ of the matrix $D$, constructed from the OM fingerprint vectors. The eigenvalues $\lambda_i$ of matrix $D$ are sorted in descending order, i.e. $\lambda_1$ is the largest eigenvalue.
The matrix elements $D_{i,j}$ equals $<V_{i}|V_{j}>$ for the atom pair $(i,j)$.\\

The number of the non-zero eigenvalues of this matrix gives the rank of the fingerprint vectors.
So the penalty function that favours one single environment for a certain element is
\begin{equation} 
\label{eq:P1}
P_1(R_1,\ldots,R_{N_{at}})=\sum_{i=2}^{N_{at}} \lambda_i=\Tr(D) -\lambda_1,
\end{equation}
In case we want to allow for up to $l$ environments, the penalty becomes 
\begin{equation} 
\label{eq:P2}
P_l(R_1,\ldots,R_{N_{at}})=\sum_{i=l+1}^{N_{at}} \lambda_i=\Tr(D) - \sum_{i=1}^{l} \lambda_i 
\end{equation}
where $R_i$ is the position of the atom $i$ in the system in Cartesian coordinates, $N_{at}$ equals the number of atoms in the system and $\Tr(D)$ is the trace of matrix D.\\
For a multi-component system, each element contributes its own penalty function and the total penalty function is the sum of all the elemental contributions.

\section{Symmetry bias derivatives}
To obtain conservative forces of the biased PES the derivative of the symmetry bias needs to be added to the physical forces. The same is true for the derivative of the biased symmetry function with respect to the lattice vectors which need to be added to the lattice derivatives in the case of PBC.
\subsection{Symmetry bias derivative with respect to atomic coordinates}
The derivative of the symmetry bias function with respect to the atomic coordinates is 
\begin{equation}
\label{eq:dP/dR}
\frac{\partial P}{\partial R}= \frac{\partial (\Tr(D)-\lambda_1)}{\partial R} =  \sum_i \left( \frac{\partial D_{i,i}}{\partial R}  \right) - \frac{\partial \lambda_1}{\partial R}.
\end{equation}
For the term $\frac{\partial \lambda_1}{\partial R}$ the Hellman Feynman Theorem is used.
$$
\frac{\partial \lambda_1}{\partial R} = \braket{X_1| \frac{\partial D}{\partial R} |X_1}
$$
with $X_1$ being the eigenvector belonging to the largest eigenvalue $\lambda_1$ of matrix $D$. The derivative of the dimensionality matrix $D$ $\frac{\partial D}{\partial R}$ depends on the derivatives of the OM fingerprints.
\begin{equation}
\label{eq:dDij/dR}
\frac{\partial D_{i,j}}{\partial R} = \sum_l \frac{\partial V_i(l)}{\partial R} V_j(l) + \frac{\partial V_j(l)}{\partial R} V_i(l)
\end{equation}
with $l$ counting over all entries in the atomic fingerprint eigenvectors $V_i$. The derivative $\frac{\partial V_i}{\partial R}$ is formed with the help of publication ~\cite{zhu2016fingerprint}.\\
Now the negative gradient of the derivative can be added to the physical forces to obtain the biased forces belonging to the biased PES.\\

\subsection{Symmetry bias derivative with respect to  lattice vectors}
Analog to the derivative with respect to the atom positions we can find the derivative with respect to the lattice vectors
\[
\frac{\partial{P}}{\partial{{\mathbf{h}}}} = \frac{\partial{ ( Tr(D) - \lambda_1 ) }}{\partial{{\mathbf{h}}}} = \sum_i \left( \frac{\partial{D_{i,i}}}{\partial{\mathbf{h}}}  \right) - \frac{\partial{\lambda_{1}}}{\partial{\mathbf{h}}}
\]
with \textbf{h} being the lattice vector matrix
\[
\mathbf{h} = \begin{bmatrix} 
h_1(1) & h_2(1) & h_3(1) \\
h_1(2) & h_2(2) & h_3(2) \\
h_1(3) & h_2(3) & h_3(3) \\
\end{bmatrix}
\]
with $h_i$ being the lattice vectors.
Like before we can use the the Hellman Feynman Theorem for the term $\frac{\partial \lambda_1}{\partial \mathbf{h}}$. This results in
$$
\frac{\partial \lambda_1}{\partial \mathbf{h}} = \braket{X_1| \frac{\partial D}{\partial \mathbf{h}} |X_1}
$$
The derivative of the matrix entries $D_{i,j}$ with respect to the lattice vectors is
\[
\frac{\partial{D_{i,j}}}{\partial{\mathbf{h}}} = \sum^{l_{fp}}_{l} \frac{\partial{V_i(l)}}{\partial{\mathbf{h}}} V_j(l) +  \frac{\partial{V_j(l)}}{\partial{\mathbf{h}}} V_i(l)
\]
where we now need the derivative of the OM fingerprints $V_i$ with respect to the lattice vectors.

To calculate the derivative of the OM fingerprints $V_k$ with respect to the lattice vectors we can apply the chain rule so that we can use the already known derivation of the OM fingerprint with respect to the atomic positions $\frac{\partial V_k}{\partial R}$. 
It is important to note that the OM fingerprints $V_k$ for atom $k$ in the system is formed by putting Gaussian type orbitals only on all atoms within a given cutoff radius around the central atom $k$ and then forming an overlap matrix from them.
Therefore, we only need to consider the atomic positions $\tilde{R}_{j}^{k}$ of all atoms $j$ in the sphere around the central atom $k$. This leads to the fact that we now have two counting schemes, one for the atoms in the sphere and one for the atoms in the main cell. 
To deal with this we introduce a function index$(i,k)$ that maps atom number $i$ from the sphere counting scheme of the central atom $k$ to the main cell counting scheme that gives back the index of the corresponding atom in the main cell counting scheme.
This results in
\[
\frac{\partial{V_{k}(l)}}{\partial{{h}}} = \sum^{N_{sp}}_{j} \frac{\partial{V_{k}(l)}}{\partial{{\tilde{R}_{j}^{k}}}} \frac{\partial{\tilde{R}_{j}^{k}}}{\partial{{h}}}
\]
with $N_{sp}$ being the number of atoms in a sphere that is formed by the cutoff radius around the central atom $k$. 
Since the derivative of the overlap matrix fingerprint is invariant under the change of the counting scheme we get
\[
\frac{\partial V_k}{\partial \tilde{R}_{j}^{k}} = \frac{\partial V_k}{\partial R_{\text{index}(j,k)}}.
\]

One needs to be aware of the fact that in periodic boundary conditions it is possible that multiple images of the same atom from the main cell can be inside the cutoff radius. 
The position of the atoms in the sphere around atom $k$ can then be described as
\[
\tilde{R}^{k}_{i} = R_{\text{index}(i,k)} + \mathbf{h} \cdot \mathbf{n}_{i,k} = \mathbf{h} \cdot R_{\text{index}(i,k)}^{\text{frac}} + \mathbf{h} \cdot \mathbf{n}_{i,k}
\]
with 
\[
\mathbf{n}_{i,k}  = \begin{bmatrix} 
a_{i,k} \\
b_{i,k} \\
c_{i,k} \\
\end{bmatrix}
\]
being the multiplier for atoms outside the periodic cell and 
\[
R_{\text{index}(i,k)}^{\text{frac}} =
\begin{bmatrix} 
\alpha_{\text{index}(i,k)} \\
\beta_{\text{index}(i,k)} \\
\gamma_{\text{index}(i,k)} \\
\end{bmatrix}
\]
the atomic coordinates in fractional form of the atom belonging to $\text{index}(i,k)$ in the main cell.

This results in
\[
 \frac{\partial{\tilde{R}_{i}^{k}}}{\partial{{h}}} = 
 \begin{bmatrix} 
\alpha_{\text{index}(i,k)} + a_{i,k} & \beta_{\text{index}(i,k)} + b_{i,k} & \gamma_{\text{index}(i,k)} + c_{i,k} \\
\alpha_{\text{index}(i,k)} + a_{i,k} & \beta_{\text{index}(i,k)} + b_{i,k} & \gamma_{\text{index}(i,k)} + c_{i,k} \\
\alpha_{\text{index}(i,k)} + a_{i,k} & \beta_{\text{index}(i,k)} + b_{i,k} & \gamma_{\text{index}(i,k)} + c_{i,k} \\
\end{bmatrix}
\]
for atom $i$ in the sphere counting scheme around the central atom $k$ in the main cell counting scheme.